\documentclass[prl,aps,twocolumn,showpacs,preprintnumbers,amsmath,amssymb]{revtex4}

\usepackage{graphicx}
\usepackage{dcolumn}
\usepackage{bm}
\usepackage{color}

\begin{document}


\title{Heavily Damped Motion of One-Dimensional Bose Gases 
in an Optical Lattice}
\author{Ippei Danshita$^{1}$}
\author{Charles W. Clark$^{2}$}
\affiliation{
{$^1$Department of Physics, Faculty of Science, Tokyo University of Science, Shinjuku-ku, Tokyo 162-8601, Japan}
\\
{$^2$Joint Quantum Institute, National Institute of Standards and Technology and University of Maryland, Gaithersburg, Maryland 20899, USA}
}

\date{\today}

\begin{abstract}
We study the dynamics of strongly correlated one-dimensional Bose gases 
in a combined harmonic and optical lattice potential subjected to sudden
displacement of the confining potential.
Using the time-evolving block decimation method, we perform a first-principles
quantum many-body simulation of the experiment of
Fertig {\it et al.} [Phys. Rev. Lett. {\bf 94}, 120403 (2005)] 
across different values of the lattice depth ranging from the superfluid to
the Mott insulator regimes.
We find good quantitative agreement with this experiment: the damping of 
the dipole oscillations is significant even for 
shallow lattices, and the motion becomes overdamped
with increasing lattice depth as observed.
We show that the transition to overdamping is attributed to the decay of 
superfluid flow accelerated by quantum fluctuations, which occurs well 
before the emergence of Mott insulator domains.
\end{abstract}

\pacs{03.75.Hh, 03.75.Lm, 05.30.Jp}
\keywords{optical lattice, superfluid-to-Mott insulator transition, , dynamical instability}
\maketitle
Systems of ultracold bosonic atoms provide new opportunities to study
many-body quantum physics in low dimensions, where thermal and quantum 
phase fluctuations play a crucial role.
The unprecedented degree of controllability attainable in cold atom experiments
has led to demonstrations of striking phenomena specific to low-dimensional 
systems, such as the purely one-dimensional (1D) Tonks-Girardeau
gases~\cite{rf:paredes,rf:kinoshita1,rf:kinoshita2} 
and the Kosterlitz-Thouless transition of 2D Bose gases~\cite{rf:hadzibabic,rf:pierre}.

An array of 1D tubes of Bose gases can be created by imposing
a strong transverse 2D optical lattice on a Bose condensed gas
trapped in a parabolic potential.
In recent experiments of this type, transport of 1D Bose gases 
has been investigated in the presence of another lattice potential along 
the 1D axis by displacing suddenly the parabolic potential~\cite{rf:stoeferle,
rf:fertig}, and by using a moving optical lattice~\cite{rf:mun}.
Strong inhibition of transport with increasing lattice depth has been observed 
in these cases.
While 1D transport has also been studied in condensed matter systems, such as 
liquid $^4{\rm He}$ absorbed in nanopores~\cite{rf:toda} and superconducting 
nanowires~\cite{rf:bezryadin}, the flexible variability of optical lattice 
parameters allows for more detailed study in ultracold atomic gases.
Moreover, the long relaxation time resulting from the diluteness of atomic
gases offers the prospect of exploring dynamical properties of the transport 
over a wide range of conditions.

Fertig {\it et al.}~\cite{rf:fertig} induced dipole 
oscillations of 1D Bose gases for different values of the axial lattice depth. 
They observed significant damping even for shallow lattices, and found 
a transition from underdamping to overdamping with increasing lattice depth.
Due to strong quantum fluctuations, mean-field theories fail even qualitatively
to describe the transport properties of 1D Bose gases, and recent theoretical 
analyses of this system have applied approaches beyond the mean-field 
approximation~\cite{rf:anatoli1,rf:ruostekoski,rf:rigol,rf:ana,rf:guido}.
Good agreement with the experiment was obtained separately for the cases of 
shallow lattices, using a truncated Wigner 
approximation~\cite{rf:ruostekoski}, and for deep lattices, 
using an extended fermionization method~\cite{rf:guido}.
However, so far the intermediate transition region remains poorly understood
because the approaches used in the previous work are no longer valid there.

In this Letter, we present a direct simulation of this experiment from the 
shallow-lattice superfluid (SF) to the deep-lattice Mott insulator (MI) 
regimes, and we identify the specific physical mechanism governing 
the transition region. 
To properly include the effects of quantum fluctuations, we use the 
time-evolving block decimation (TEBD) method, which provides precise ground 
states and real-time evolution of 1D quantum lattice systems~\cite{rf:vidal1}.
We show that the transition to overdamping is a manifestation of 
the decay of superfluid flow accelerated by strong quantum fluctuations, 
rather than that of the emergence of the MI domains.

We begin with the Bose-Hubbard model~\cite{rf:fisher}, 
\begin{eqnarray}
H &=& -J\sum_{j}(\hat{b}^{\dagger}_j \hat{b}_{j+1} + {\rm h.c.}) + \frac{U}{2}\sum_{j}\hat{n}_j(\hat{n}_j-1)
   \nonumber\\
   & & + \Omega \sum_{j}(j+X_{\rm c}/d)^2 \hat{n}_j,
\label{eq:BHH}
\end{eqnarray}
where $\hat{b}^{\dagger}_j$ creates a boson 
on the $j$-th site, and $\hat{n}_j$ is the number operator.
The parameters $J$, $U$, and $\Omega$ are related to independently-measurable 
experimental quantities.
The hopping energy, $J$, is expressed as
%
$
J=A\left(V_0/E_{\rm R}\right)^B
                    {\rm exp}\left(-C\sqrt{V_0/E_{\rm R}}\right)E_{\rm R},
$
%
where $A=1.397$, $B=1.051$, and $C=2.121$~\cite{rf:ana}.
$V_0$ is the axial lattice depth and $E_{\rm R} = h^2/(8 m d^2)$ is the recoil energy, 
where $m$ is the atomic mass and $d$ is the lattice spacing.
The onsite interaction, $U$, is given by
%
$
U = 2 (a_s/d) \sqrt{2\pi V_{\perp}/E_{\rm R}}
                  \left(V_0/E_{\rm R}\right)^{1/4}E_{\rm R},
$
%
where $a_s$ is the $s$-wave scattering length and $V_{\perp}$ is the depth of
the lattice in the transverse directions.
The parabolic trap is centered at $-X_{\rm c}$, and its curvature, $\Omega$, is 
given by
%
$
\Omega=m\omega_{\rm T}^2 d^2/2
$,
%
where $\omega_{\rm T}$ is the frequency of the external harmonic potential.
In the experiment, $X_{\rm c}=0$ for $t<0$, and it is suddenly displaced 
to $X_{\rm c}=x_0$ at $t=0$ to cause dipole motions of the Bose gas. 
Note that the Bose-Hubbard model is quantitatively valid only when the 
lattice depth is sufficiently large as $V_0 \gtrsim 2 E_{\rm R}$~\cite{rf:ana}.

To compare our calculations with the experiment of Fertig 
{\it et al.}~\cite{rf:fertig}, we use $a_s=5.31 \, {\rm nm}$,
$d=405 \, {\rm nm}$, $V_{\perp} = 30 E_{\rm R}$, and 
$\omega_{\rm T}=2\pi \times 60 \, {\rm Hz}$.
In the experiment, the atoms were confined in a 2D array of decoupled
tight 1D tubes and an additional periodic potential was added along the
direction of the tubes.
We focus on the central tube with the number of atoms $N=81$.

To treat the dynamics at zero temperature associated with Eq.~(\ref{eq:BHH}), 
we use the TEBD method, which allows us to compute 
accurately the evolution of many-body wave functions of 1D quantum lattice 
systems~\cite{rf:vidal1}.
We choose the maximum number of atoms per site 
$n_{\rm max}=5$ and retain up to $\chi = 200$ states in the adaptively
selected Hilbert space 
(for the definition of $\chi$ and a detailed convergence 
analysis, see Ref.~\cite{rf:EPAPS}).
Setting $X_{\rm c}=0$, we first calculate the ground states of 
Eq.~(\ref{eq:BHH}) via propagation in imaginary time.
At $t=0$ the trap center $X_{\rm c}$ is suddenly displaced by the distance 
$x_0$ to cause dipole motions, and we simulate the dynamics of the system
via propagation in real time.
In the experiment, the initial displacement was fixed to be 
$x_0\simeq 8d$ 
and $V_0/E_{\rm R}$ was changed in the range
of $0\leq V_0/E_{\rm R} \leq 9$~\cite{rf:fertig}.  
Here we also study the dependence of the damping on $x_0$, so as to 
reveal the effects of the decay of superfluid flow.

\begin{figure}[b]
\includegraphics[scale=0.42]{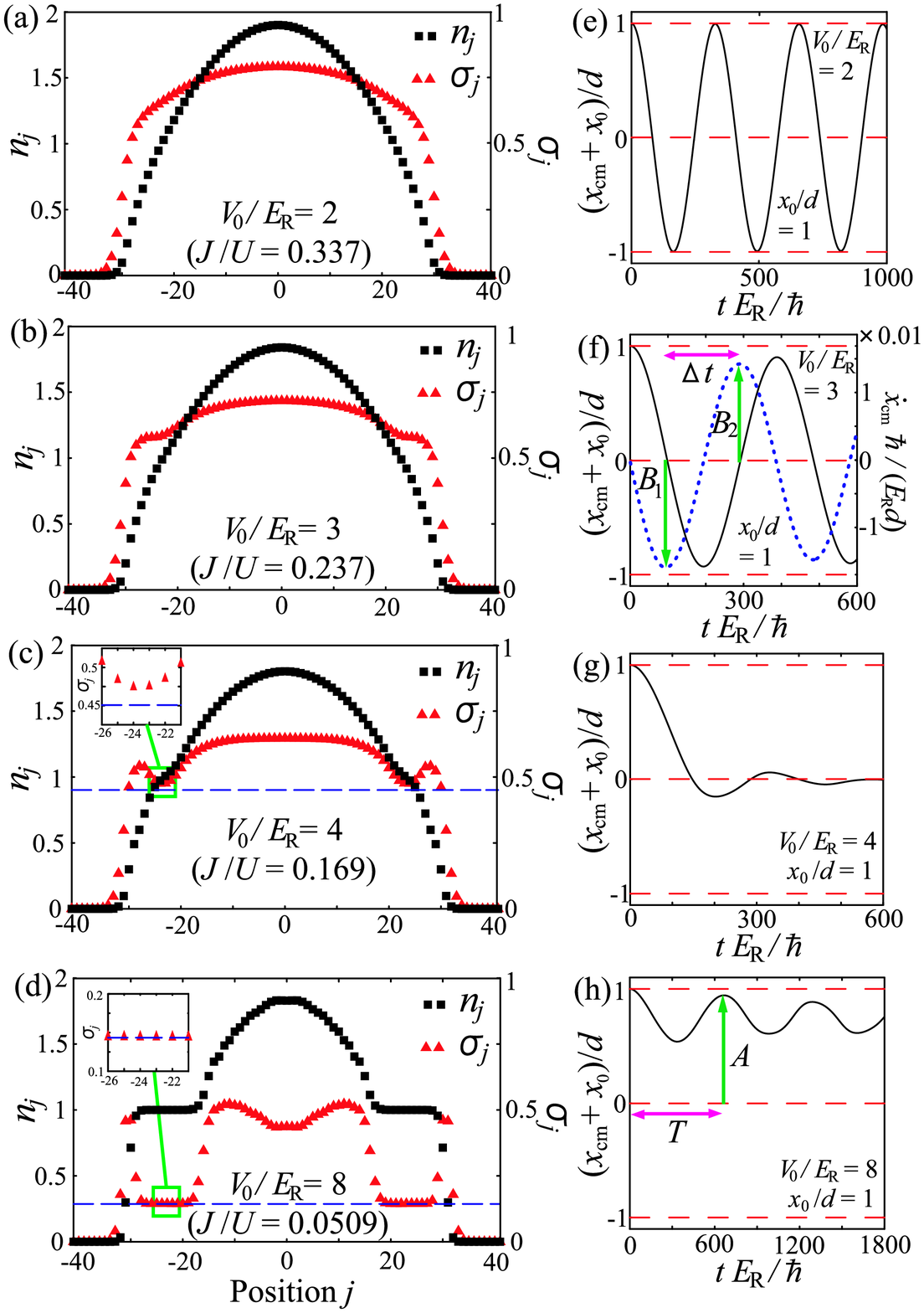}
\caption{\label{fig:GSandCM}
(color online)
Results at four values of $V_0/E_{\rm R}$: 2 (a) and (e), 3 (b) and (f), 
4 (c) and (g), and 8 (d) and (h).
In the left four figures, we show the local densities 
$n_j\equiv \langle \hat{n}_j \rangle$ (squares) and fluctuations 
$\sigma_j$ (triangles) of the ground states.
The dashed lines in (c) and (d) represent the fluctuations 
for the unit-filling in the homogeneous Bose-Hubbard system, calculated with 
the use of the infinite-size version of the TEBD~\cite{rf:vidal2}. 
In the right four figures, we show the time evolution of the c.m.~position
$x_{\rm cm}(t)$ for a small displacement $x_0=d$.
In (f), $\dot{x}_{\rm cm}(t)$ is also plotted by a dotted line.
}
\end{figure}

In Figs.~\ref{fig:GSandCM}(a)-(d), we show the local densities 
$n_j\equiv \langle \hat{n}_j \rangle$, and their fluctuations, 
$\sigma_j\equiv (\langle \hat{n}_j^2 \rangle - n_j^2)^{1/2}$, of 
the ground states of the system for lattice depths $V_0/E_{\rm R}=2$, $3$, 
$4$, and $8$. 
When $V_0 = 2 E_{\rm R}$, the density profile is smooth, with the reflected
parabolic shape of the confining potential that is characteristic of SF phases. 
As $V_0/E_{\rm R}$ is increased, the density profile around $n_j=1$ becomes
distorted 
and the fluctuation $\sigma_j$ starts to take minima at $n_j =1$
as seen in Fig.~\ref{fig:GSandCM}(c), preluding 
the emergence of the incompressible MI domains with unit filling.
When $V_0=8E_{\rm R}$ (Fig.~\ref{fig:GSandCM}(d)),
the unit-filling MI plateaus are formed and its presence is convinced by 
the fact that the value of $\sigma_j$ in the plateaus coincides with the
value of the fluctuation in the homogeneous system for the unit filling and
the same value of $J/U$~\cite{rf:rigol2}.

In Figs.~\ref{fig:GSandCM}(e)-(h), we show the c.m.~position 
$x_{\rm cm}=N^{-1}d\sum_j j \langle \hat{n}_j \rangle$ as a function of time
for $x_0=d$ and different values of $V_0/E_{\rm R}$.
Note that this displacement amplitude is substantially smaller than that 
employed in the experiment of Ref.~\cite{rf:fertig}; we return to the
specific experimental simulation below, after identifying an important point of
physics that is seen most clearly in the small-displacement regime. 
The c.m.~velocity $\dot{x}_{\rm cm}(t)$
for $V_0=3 E_{\rm R}$ is also shown by a dotted line 
in Fig.~\ref{fig:GSandCM}(f).
When $V_0=2 E_{\rm R}$, the oscillation is hardly damped; this means that
the Bose gas behaves as a superfluid as long as its velocity is small.
As $V_0/E_{\rm R}$ is increased, noticeable damping is already seen at
$V_0=3E_{\rm R}$.
At $V_0=4E_{\rm R}$, the oscillation is drastically damped 
while it remains in the underdamped regime according to the 
definition of underdamping mentioned below.
This means that the suppression of fluctuations at $n_j=1$
results in significant reduction of the mobility of the Bose gas even before
the formation of the MI domains.
When the lattice is very deep ($V_0=8E_{\rm R}$), the motion is overdamped.

\begin{figure}[tb]
\includegraphics[scale=0.7]{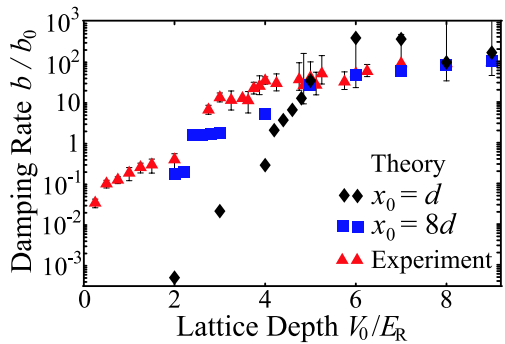}
\caption{\label{fig:dampingrate1}
(color online)
Damping rate $b/b_0$ versus $V_0/E_{\rm R}$ for $x_0=d$ (diamonds) and 
$x_0 = 8 d$ (squares).
The triangles are the experimental results of Ref.~\cite{rf:fertig}.
}
\end{figure}

Following the method used by Ref.~\cite{rf:fertig}, 
we model the c.m.~motion of the Bose gas in our results as a damped harmonic oscillator,
%
$
m^{\ast}\ddot{x}=-b\dot{x}-k x,
\label{eq:dampedharmonic}
$
%
to extract a damping rate $b/b_0$ as a function of $V_0/E_{\rm R}$ and $x_0/d$,
where $m^{\ast}$ is the effective mass, $k\equiv m\omega_{\rm T}^2$, and 
$b_0\equiv 2m\omega_{\rm T}$~\cite{rf:footnote2}.
We regard the c.m.~motion as underdamped if the c.m.~position 
oscillates across the trap center, $x=-x_0$.
In this definition, for instance, the c.m.~motions for 
$V_0/E_{\rm R} = 2,3$, and $4$ and $x_0=d$ shown in Figs.~\ref{fig:GSandCM}(e)-(g)
are underdamped.
On the other hand, the c.m.~motion is regarded as overdamped if the first minimum
of the c.m.~position does not reach the trap center as seen in 
Fig.~\ref{fig:GSandCM}(h).
To obtain the damping rate, we use the minimum and maximum values of 
$x_{\rm cm}$ and $\dot{x}_{\rm cm}$ within the first oscillation.
For underdamped motion, where $b<b_0$, the damping rate is given
by $b=2m^{\ast}{\rm ln}(B_1/B_2)/\Delta t$, where $B_1$ and $B_2$ 
are the absolute values of $\dot{x}_{\rm cm}$ at the first and second extrema, 
respectively, and $\Delta t$ is the time interval between the two extrema 
as indicated in Fig.~\ref{fig:GSandCM}(f).
For overdamped motion, where $b>b_0$, the damping rate is determined by
solving $\bar{x}_{\rm cm}(T)=A$ for $b$, where $A$ and $T$ are 
the values of $x_{\rm cm}$ and $t$ at the first maximum 
(see Fig.~\ref{fig:GSandCM}(h)), and $\bar{x}_{\rm cm}(t)$ is the solution to 
the overdamped harmonic oscillator equation.

In Fig.~\ref{fig:dampingrate1}, we show the damping rates
for $x_0=d$ and $8d$ as a function of $V_0/E_{\rm R}$, together with the 
experimental data of Ref.~\cite{rf:fertig} where $x_0\simeq 8d$.
In the experiment, the motion becomes overdamped with increasing the lattice 
depth, starting between $V_0/E_{\rm R}=2$ and $3$.
Our TEBD simulations for $x_0=8d$ are in good agreement with the experimental 
results; the onset of the overdamping lies between $V_0/E_{\rm R}=2.2$ and 
$2.4$, where the MI domains are not present.
This reveals that the transition to overdamping observed in the experiment
is not due to the emergence of the MI domains.
Notice that although the damping rate $b/b_0$ exhibits a jump 
at the transition point, the actual mobility of the Bose gas is gradually 
reduced and this jump is only superficial~\cite{rf:EPAPS}.

\begin{figure}[tb]
\includegraphics[scale=0.65]{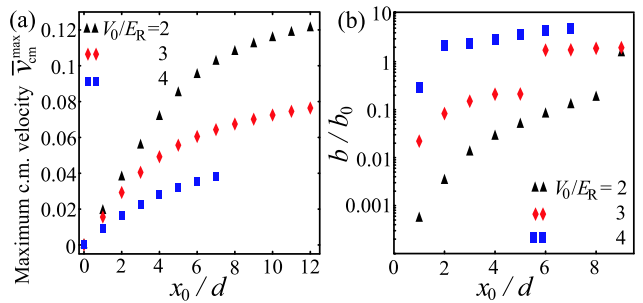}
\caption{\label{fig:dampvelofdisp}
(color online)
Maximum c.m.~velocities $\bar{v}_{\rm cm}^{\rm max}\equiv |B_1|$ (a) and 
damping rates $b/b_0$ (b) for $V_0/E_{\rm R}=2$ (triangles), $3$ (diamonds), 
and $4$ (squares) as functions of $x_0/d$.
}
\end{figure}
The mechanism of damping in the transition region is suggested in 
Fig.~\ref{fig:dampvelofdisp}, which shows the maximum c.m.~velocities
$\bar{v}_{\rm cm}^{\rm max}$ and  
the damping rates for different values of $V_0/E_{\rm R}$
as functions of $x_0/d$.
For a given value of the lattice depth, the damping becomes 
more significant as $\bar{v}_{\rm cm}^{\rm max}$ increases, and the motion 
eventually changes from underdamping to overdamping.
This behavior is analogous to the breakdown of superfluid flow 
observed in the experiment of Ref.~\cite{rf:mun} by using a moving optical
lattice, where a transition from superfluid flow to dissipative flow is induced
as the lattice velocity is increased~\cite{rf:footnote3}.
Hence, this strong velocity dependence of the damping rate implies that the transition
to overdamping can be attributed to the breakdown of superfluid flow.

\begin{figure}[tb]
\includegraphics[scale=0.45]{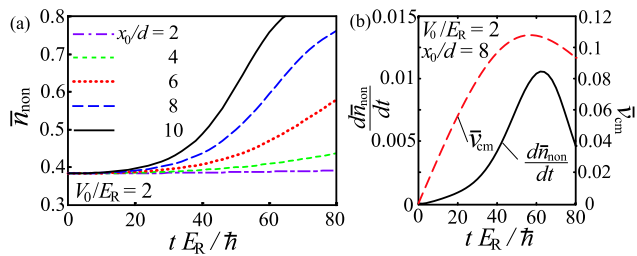}
\caption{\label{fig:depgrowth}
(color online)
(a) Time evolution of the non-condensate fraction $\bar{n}_{\rm non}(t)$ 
at $V_0 = 2 E_{\rm R}$ for different values of the displacement.
(b) Time evolution of 
$d\bar{n}_{\rm non}/dt$ (solid line) and 
$\bar{v}_{\rm cm}$
(dashed line) for $V_0= 2 E_{\rm R}$ and $x_0 = 8 d$.
}
\end{figure}

\begin{figure}[tb]
\includegraphics[scale=0.7]{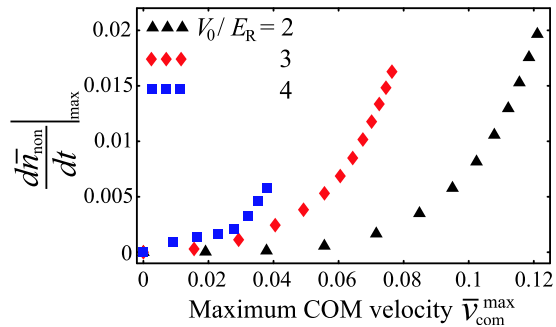}
\caption{\label{fig:growthrate}
(color online) Maximum growth rates $d\hat{n}_{\rm non}/dt|_{\rm max}$ of
the non-condensate fraction for $V_0/E_{\rm R}= 2$ (triangles), $3$ (diamonds),
and $4$ (squares) as functions of $\bar{v}_{\rm cm}^{\rm max}$.
}
\end{figure}

To confirm this suggestion, we show the time evolution of the 
non-condensate fraction $\bar{n}_{\rm non}(t) \equiv (N-N_{\rm c})/N$ 
in Fig.~\ref{fig:depgrowth}(a), where the number of the condensate bosons
$N_{\rm c}$ is defined as the largest eigenvalue of the one-body density
matrix $\langle \hat{b}_j^{\dagger} \hat{b}_l \rangle$~\cite{rf:penrose}.
Notice that despite the one-dimensionality of our system, the condensate 
fraction $N_{\rm c}/N$ of the superfluid ground states is of order one because 
our system is finite.
Since as a consequence of the breakdown of superfluid flow,
excitations kick particles out of the condensate increasing the non-condensate
fraction,
the breakdown can be identified by growth of
the non-condensate fraction in time as done in the experiment of 
Ref.~\cite{rf:mun}.
In Fig.~\ref{fig:depgrowth}(a), for small displacement ($x_0=2d$)
$\bar{n}_{\rm non}$ hardly changes as $t$ increases; accordingly the superfluid
flow is stable. 
The growth of $\bar{n}_{\rm non}$ in time becomes more prominent as $x_0/d$ 
increases, namely, as the damping becomes more significant.
This means that the c.m.~motion is damped by the transference of kinetic energy
into non-condensate excitations accompanied by the breakdown of 
superfluid flow.
In Fig.~\ref{fig:depgrowth}(b), one sees that the growth rate 
$d\bar{n}_{\rm non}/dt$ takes its maximum value immediately after 
the c.m.~velocity peaks.
In Fig.~\ref{fig:growthrate}, we show the maximum value of the growth rate
$d\bar{n}_{\rm non}/dt|_{\rm max}$ for different values of the lattice depth
as a function of $\bar{v}_{\rm cm}^{\rm max}$.
The maximum growth rate is increased 
as either $\bar{v}_{\rm cm}^{\rm max}$ or $V_0/E_{\rm R}$ is increased.
When $x_0=8d$ as chosen in the experiment~\cite{rf:fertig}, 
the breakdown of superfluid flow is so severe that it leads to significant 
damping already in the underdamped regime. 
As $V_0/E_{\rm R}$ is increased, superfluid flow breaks down 
even more severely and the c.m.~motion gradually changes to overdamping.
Therefore, the transition to overdamping observed in the experiment is due to 
the breakdown of superfluid flow.
One can experimentally corroborate this interpretation by measuring 
the non-condensate fraction and specifying its growth associated with the significant 
damping because the non-condensate fraction is one of the typical observables in 
experiments~\cite{rf:mun}.

We now discuss the effects of quantum fluctuations on the breakdown of 
superfluid flow.
As seen in Fig.~\ref{fig:growthrate}, in the 1D Bose gases 
$d\bar{n}_{\rm non}/dt|_{\rm max}$ grows gradually with increasing
$\bar{v}_{\rm cm}^{\rm max}$ in contrast to 3D Bose condensates,
where a transition from superfluid to dissipative flow occurs
very sharply~\cite{rf:mun}.
Moreover, the superfluid flow breaks down significantly even at much lower
velocity than the critical velocity $v_{\rm cr}$ for the dynamical
instability predicted by mean-field theories~\cite{rf:smerzi} (for instance,
$v_{\rm cr}\simeq 0.288 E_{\rm R}d/\hbar$ when 
$V_0=2E_{\rm R}$~\cite{rf:landauI}).
The smearing and reduction of the critical velocity stem from strong quantum 
fluctuations specific to the 1D systems~\cite{rf:mun}, which allow for the 
decay of the supercurrent through macroscopic quantum 
tunneling~\cite{rf:anatoli2}.
Thus, the quantum fluctuations accelerate the decay of 
superfluid flow, leading to the heavily damped motion observed in the 
experiment.


In conclusion, we have studied damped dipole oscillations of one-dimensional
Bose gases in an optical lattice by using the time-evolving block decimation
method.
We have shown that the center of mass motion changes from underdamping 
to overdamping as observed in the experiment~\cite{rf:fertig} because 
strong quantum fluctuations accelerate the decay of superfluid flow, which is 
characterized by prominent growth of the non-condensate fraction in time.

\begin{acknowledgments}
The authors thank Trey Porto for helpful discussions and for providing the
experimental data.
I. D. acknowledges discussions with S. Konabe, 
S. Tsuchiya, and K. Kamide and support from a Grant-in-Aid from JSPS.
This work was partially supported by the NSF 
under Physics Frontiers Center award PHY-0822671.
\end{acknowledgments}


\begin{thebibliography}{99}
\bibitem{rf:footnote2}
As seen in Fig.~\ref{fig:GSandCM}(h), the motion in the overdamping
region still oscillates and is different from an overdamped 
oscillator represented by a simple mechanical model.
However, we model this motion as an overdamped oscillator to compare
our results with the experiment~\cite{rf:fertig}.
\bibitem{rf:footnote3}
A widely used definition of 
superfluid flow is ``flow without dissipation",
which we also use in this paper.
\bibitem{rf:landauI}
Mean-field theories also predict the Landau instability characterized by 
excitations with negative energies~\cite{rf:smerzi}.
However, it cannot destabilize superfluid flow
at zero temperature~\cite{rf:sarlo}, and is not relevant to our case.
\end{thebibliography}
\end{document}